\documentclass[sn-standardnature,iicol]{sn-jnl}

\usepackage{color, colortbl}
\definecolor{Gray}{gray}{0.9}

\usepackage{lineno}

\ifpdf
    \graphicspath{{Figs/}}
\else
    \graphicspath{{Figs/}}
\fi

\jyear{2023}%

\theoremstyle{thmstyleone}%
\theoremstyle{thmstyletwo}%

\theoremstyle{thmstylethree}%

\PassOptionsToPackage{normalem}{ulem}
\usepackage{ulem}
\providecolor{added}{rgb}{1,0,0}
\providecolor{deleted}{rgb}{1,0,0}
\newcommand{\added}[1]{#1}
\newcommand{\deleted}[1]{}

\usepackage{setspace}

\begin{document}

\title[ ]{A Hybrid Integrated Quantum Key Distribution Transceiver Chip}

\author*[1,2]{\fnm{Joseph A.} \sur{Dolphin}}\email{joseph.dolphin@toshiba.eu}

\author[1]{\fnm{Taofiq K.} \sur{Paraiso}}

\author[1]{\fnm{Han} \sur{Du}}

\author[1]{\fnm{Robert I.} \sur{Woodward}}

\author[1]{\fnm{Davide G.} \sur{Marangon}}

\author[1]{\fnm{Andrew J.} \sur{Shields}}

\affil[1]{\orgname{Toshiba Europe Ltd}, \orgaddress{\street{208 Cambridge Science Park Milton Rd}, \city{Cambridge}, \postcode{CB4 0GZ}, \country{UK}}}

\affil[2]{\orgdiv{Department of Engineering}, \orgname{University of Cambridge}, \orgaddress{\street{Trumpington St}, \city{Cambridge}, \postcode{CB2 1PZ}, \country{UK}}}


\abstract{
Quantum photonic technologies, such as quantum key distribution, are already benefiting greatly from the rise of integrated photonics. However, the flexibility in design of these systems is often restricted by the properties of the integration material platforms. Here, we overcome this choice by using hybrid integration of ultra-low-loss silicon nitride waveguides with indium phosphide electro-optic modulators to produce high-performance quantum key distribution transceiver chips. Access to the best properties of both materials allows us to achieve active encoding and decoding of photonic qubits on-chip at GHz speeds and with sub-1$\%$ quantum bit error rates over long fibre distances. We demonstrate bidirectional secure bit rates of 1.82 Mbps over 10 dB channel attenuation and positive secure key rates out to 250 km of fibre. The results support the imminent utility of hybrid integration for quantum photonic circuits and the wider field of photonics.}




\maketitle

\section{Introduction}\label{sec1}

Methods of data encryption are in constant competition with the abilities of attackers hoping to decrypt them. In an age of ever-increasing data exchange and the rising threat from quantum computation, quantum key distribution (QKD) provides a method of symmetric key exchange based on the fundamental principles of quantum mechanics \citep{Gisin.2002,Xu.2020}. By encoding quantum information into the states of single photons, a secure key can be exchanged between two parties with information-theoretic security. Over the last 40 years this technology has matured from a theoretical concept to field trials and commercial products \citep{Peev.2009,Sasaki.2011,Dynes.2019,Chen.2021}. However, the hardware required for QKD remains bulky and expensive \citep{Diamanti.2016}, which threatens to restrict QKD to only the most cost-insensitive of applications. 

It has thus been recognised that a move towards integrated photonics is an appealing option for QKD. Miniaturisation of the optical components into photonic integrated circuits (PICs) promises to increase the manufacturability of QKD systems to a point where widespread adoption becomes more feasible. In recent years, these 'on-chip' QKD systems have been demonstrated in a wide variety of formats, such as time-bin encoding \citep{Tanaka.2012,Sibson.2017,Sibson.2017b,Geng.2019,Paraiso.2019,Semenenko.2020,Paraiso.2021,Beutel.2021,Sax.2023}, polarisation encoding \citep{Ma.2016,Bunandar.2018,Cao.2020,Wei.2020,Zhu.2022,Avesani.2021,Du.2023}, free-space channel \citep{Avesani.2021}, measurement-device independent QKD \citep{Cao.2020,Semenenko.2020,Wei.2020} and continuous-variable QKD \citep{Zhang.2019}. Recent efforts have achieved milestones such as integrated detectors \citep{Beutel.2021}, a complete real time chip-based QKD system based on pluggable modules \citep{Paraiso.2021}, combined electronic and photonic integration \citep{Zhu.2022,Sax.2023}\added{, and an integrated transmitter used in the demonstration of the highest QKD secure key rate to date \citep{Li.2023}.}

However, the performance of QKD-PICs is constrained by the properties of the material platforms currently available, particularly in the case of the quantum receiver circuits. Here, any optical losses directly subtract from final secure key rates, making receiver loss a key performance indicator for the overall system.  Low-loss receiver circuits can be produced from materials such as silica (SiO\textsubscript{2}), silicon nitride (SiN) or silicon oxynitride (SiO\textsubscript{x}N\textsubscript{y}), but these platforms lack high-speed modulators. Receiver circuits without high-speed modulators must rely on passive structures to measure orthogonal measurement bases. This typically comes with downsides, such as the need for increased detector numbers or the need to use higher degrees of freedom, such as time-of-arrival. Indium phosphide (InP), a commonly used PIC platform with effective high-speed electro-optic phase modulators (EOPMs), suffers from relatively high waveguide propagation losses (1 - 4 dBcm\textsuperscript{-1}) \citep{Smit.2014} which discourages its use in receivers. The only platform combining low propagation losses with high-speed modulation is silicon photonics. However, a weak electro-optic effect has meant that for high-speed modulation silicon photonics usually relies on plasma dispersion effects, such as carrier depletion modulators (CDMs). In recent years work on CDMs has increased the achievable bandwidths above 50 GHz \citep{Zhou.2020}, but they still exhibit comparatively poor performance with regards to baseline loss, phase-dependent loss \citep{Sibson.2017b} and phase-modulation efficiency when\deleted{ to} compared to EOPMs \citep{Witzens.2018,Sinatkas.2021}. To-date, the majority of on-chip QKD demonstrations have opted for passive receivers, with no examples of active on-chip receivers operating above 10 MHz. 

As a solution to the limitations of the available PIC platforms we identify hybrid integration, where two different materials are combined in one photonic circuit. Others have already identified hybrid integration as an important next step for quantum integrated photonics \citep{Elshaari.2020,Wang.2021}. Indeed, for classical integrated photonics, hybrid integration has already emerged as an important capability. This has usually come in the context of integrating III/V lasers onto silicon photonics, which does not possess a native lasing ability. A number of different hybrid integration approaches have been developed, including flip-chip\citep{Theurer.2020}, edge-coupling (also called butt-coupling) \citep{Boust.2020}, and photonic wire bonding \citep{Billah.2018}, as well as heterogeneous methods \citep{Jones.2019,Snigirev.2023}. However, to date the application of hybrid integration to quantum integrated photonics has been notably limited \citep{Agnesi.2019,Li.2022}, with no full demonstrations of QKD.

In this work, we present the results from a prototype edge-coupled hybrid QKD transceiver. By combining an ultra-low-loss SiN interferometer with the superior modulation characteristics of an InP EOPM, we demonstrate actively modulated transmitter and receiver circuits with low insertion loss ($<$8 dB), low operating voltage (4 V) and high clock rate (1 GHz). We measure reliably $<$ 1 dB interface loss between the SiN and InP PICs, a value that compares well to other hybrid integration approaches and supports the use of edge-coupling when low insertion loss is important. In order to benchmark performance, we exchange quantum signals between two of our transceivers and thus estimate achievable secure key rates for a complete real time QKD system. We demonstrate simultaneous and bidirectional operation, achieving 1.82 Mbps secure key rate over a 10 dB channel loss. Further, we test our system over real fibre spools, recording positive secure key rates over 250 km of low-loss fibre in the finite key regime. Long-term testing of the system is used to demonstrate the intrinsic stability of the PIC platform. These results underline the potential of hybrid integration to expand the capabilities of QKD-PICs, as well as the wider field of quantum integrated photonics. 


\section{Results}\label{sec2}

\subsection{Hybrid Integrated Transceivers}

\begin{figure*}[h]
\centering    
\includegraphics[width=1.0\textwidth]{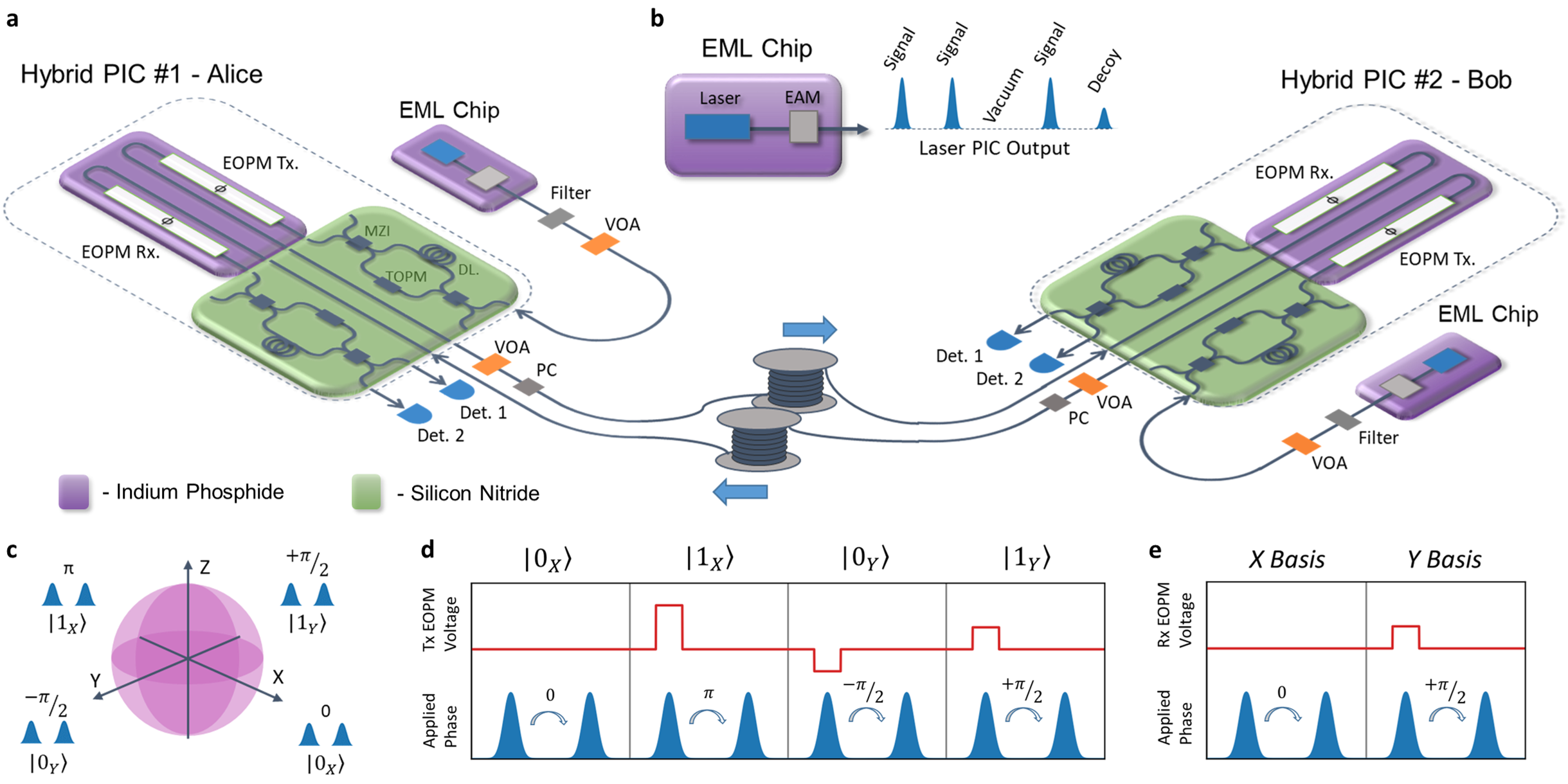}
\caption[SKR]{An overview of the hybrid transceiver experiment. \textbf{(a)} A schematic diagram of the experimental setup including hybrid chip circuits. Two hybrid PICs are shown connected via a duplex fibre channel of either emulated loss or real fibre spools. The visible on-chip components include EOPMs, mach-Zehnder interferometers (MZIs), TOPMs and delay lines (DLs). Variable optical attenuators (VOAs) are used to bring the photon flux down to single photon levels. Polarisation controllers (PC) are used to align the axis of polarisation between PICs. Each hybrid PIC is connected to an externally modulated laser (EML) chip and two single photon detectors. \textbf{(b)} Example output from the EML chips. Short gain-switched pulses are variably attenuated to produce three different intensities; signal, decoy and vacuum. \textbf{(c)} The four quantum states utilised by the system, as projected onto the time-bin Bloch sphere. \textbf{(d)} The operating principle of the transmitter. On top of the figure is the required phase states. The red line shows the applied Tx EOPM signal, whilst the blue pulses show the resulting phase relationship applied. \textbf{(e)} The operating principle of the receiver. The red line shows the applied Rx EOPM signal, whilst the blue pulses show the resulting phase relationship applied, which thus determines the measurement basis.}
\label{fig:Schematic}
\end{figure*}

A schematic containing two of the hybrid QKD transceiver chips is shown in Figure 1.a. The hybrid chip consists of two parts, a SiN asymmetric Mach-Zehnder interferometer (AMZI) and an InP EOPM, permanently bonded via an edge-coupled interface. \added{Efficient coupling of light across the interface requires three things: smooth facets, mode-size matching and precise alignment. In our devices, spot-size converter structures on both PICs provide the mode-size matching (both to 10$\mu$m mode field diameter) and also increase the tolerance to misalignment. Index-matched epoxy was used for bonding. Transmitted optical power through multiple interface structures is monitored throughout the bonding process to ensure that alignment is maintained in the various degrees of freedom until the adhesive is set.}

Each hybrid chip contains two entirely separate but identical optical circuits which are used as transmitter and receiver circuits respectively. Both laser pulse generation and photon detection are performed externally. Each assembled hybrid PIC occupies a footprint of 22 x 10 mm. 

\begin{figure*}
\centering    
\includegraphics[width=0.7\textwidth]{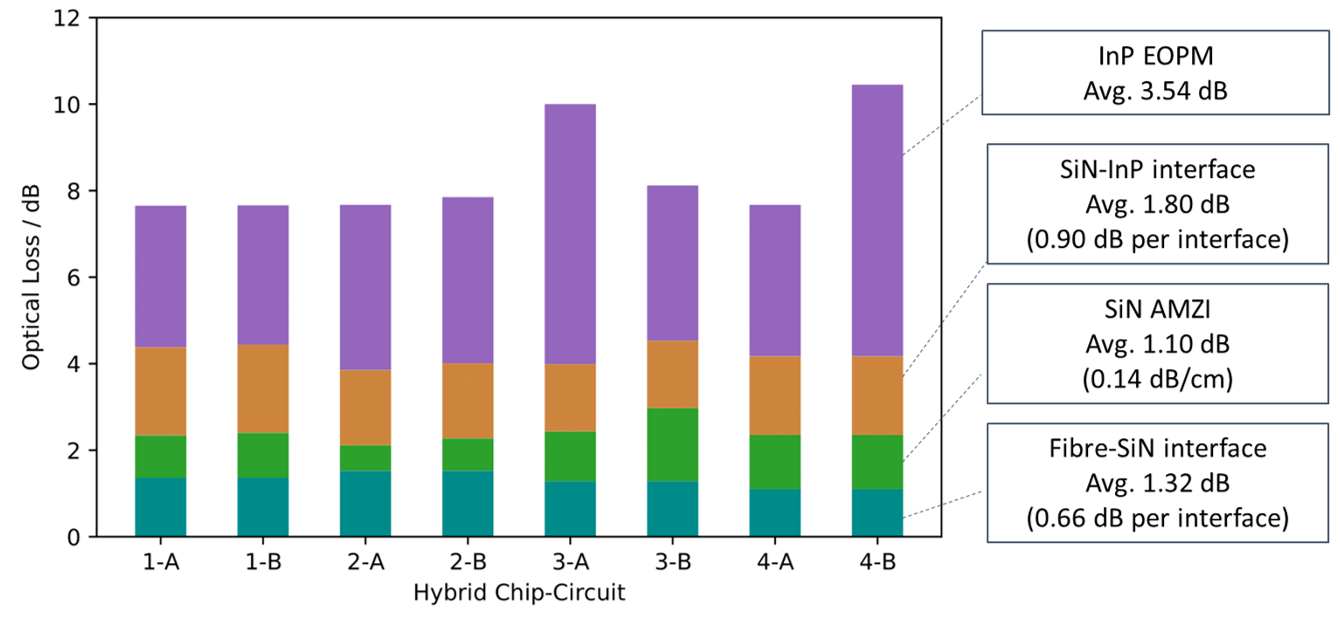}
\caption[SKR]{Breakdown of optical losses on our four manufactured hybrid photonic circuits. Loss contributions for each PIC have been estimated from various characterisation measurements. Average values and further information are given to the right. Total losses represent a complete loss from input fibre to output fibre.}
\label{fig:Loss}
\end{figure*}

The operating principle of the hybrid chip is as follows. The external laser source supplies a steady train of phase-randomised weak coherent pulses at a 1 GHz repetition rate. The AMZI structure splits these pulses into early and late time-bins with equal amplitude and separated by 500 ps. The transmitter circuit operates using four quantum states encoded by the relative phase of these early and late time-bin pulses. The four states form two mutually unbiased measurement bases, depicted in Figure 1.c and herein referred to as X and Y. After the AMZI, the pulse pairs proceed onto the transmitter (Tx) EOPM, wherein states are encoded by phase modulation of the early time-bin pulses only, as depicted by Figure 1.d. Symmetrically, in the receiving circuit, the receiver (Rx) EOPM is first used to modulate the early time-bin pulses and thus rotate the measurement basis as required, depicted in Figure 1.e. The pulse pairs then proceed to the receiving AMZI where the interference of early and late time-bins results in output at one of two external detectors ports. Thermo-optic phase modulators (TOPMs) within the SiN PICs allow the setup of the AMZIs to be finely adjusted. The high bandwidth of the EOPMs enables the early pulses to be phase-modulated with minimal impact on the late pulses. 

In total four hybrid chips were assembled, each with two independent optical circuits. The total optical loss is a key performance factor for the receiver and so losses were characterised for each circuit. Redundant waveguide structures within the PICs allowed us investigate the fibre-SiN and SiN-InP interface losses. The results are shown in Figure 2. The InP EOPMs were the most significant source of loss, typically around 3.5 dB. We measure an average facet loss of 0.66 dB for the fibre-SiN interfaces and 0.90 dB for the SiN-InP interfaces. For the SiN delay lines, we deduce an average propagation loss of 0.14 dBcm\textsuperscript{-1}, including bending losses associated with the spiral structure. We note that improved manufacturing has reduced SiN propagation losses such that they are now the least significant contribution to circuit loss. In total, 6 out of 8 hybrid circuits showed total fibre-to-fibre losses between 7.6 and 8.1 dB, with two outliers around 10 dB due to elevated InP loss. These results are an encouraging indication of the reproducible low loss of the edge-coupling approach. Further details of the optical loss measurements are given in Methods \ref{OLM}. Chips \#1 and \#2 were chosen for the further experiments described herein.

\subsection{Bidirectional QKD}

\begin{figure*}
\centering    
\includegraphics[width=0.7\textwidth]{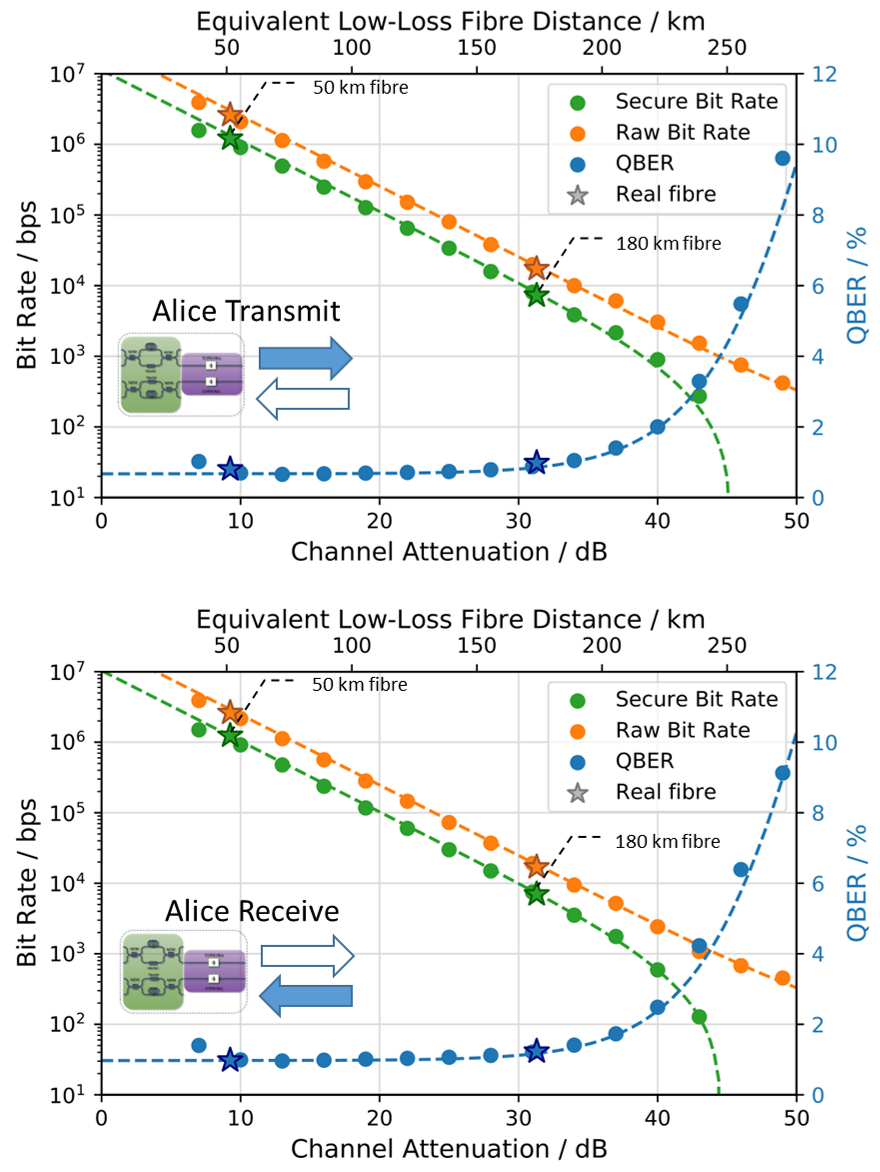}
\caption[SKR]{Achieved bidirectional and simultaneous quantum key distribution performance. Raw bit rate, secure bit rate and QBER are given for the Alice transmit (forwards) and Alice receive (backwards) directions. Data has been taken at a range of channel attenuations and for 50 km and 180 km of low-loss fibre. The dotted lines show the results of simulations based on measured system parameters.}
\label{fig:SKR}
\end{figure*}

Each of our hybrid chips is capable of simultaneously transmitting and receiving quantum signals. To illustrate this capability, we operated two of identical chips with one another bidirectionally. Bidirectional operation readily yields doubled secure key rates compared to a single channel, but could also more importantly prove useful for more complex network topologies. As shown in Figure 1.a, two hybrid PIC assemblies were connected with a duplex single-mode fibre channel. VOAs on the channels were used to simulate fibre loss or could be replaced with fibre spools for real fibre testing. One EML chip for each hybrid produced a pulse train imprinted with a three-level decoy-state \citep{Lo.2005} intensity pattern, as depicted by Figure 1.b. Detection was performed by external superconducting nano-wire single photon detectors (SNSPDs) running at 85\% efficiency and $\sim$100 dark counts per second. All high-speed on-chip components were controlled using the synchronised outputs of two arbitrary waveform generators (AWGs). We found the Tx EOPMs to require a modulation (V\textsubscript{$\pi$}) of $\sim$4 V and the Rx EOPM to require a $\pi/2$ rotation of $\sim$2 V. Both hybrid PICs were mounted on Peltier elements connected to thermo-electric controllers in order to maintain stable chip temperatures at 23$^{\circ}$C with an accuracy of \deleted{$0.01^{\circ}$C}\added{$0.001^{\circ}$C}.

For these experiments we used an optimised version of the three-level decoy-state BB84 protocol \citep{Wang.2005,Lo.2005}. Basis selections for both Alice and Bob were biased towards the X basis, with a constant ratio of 15:1. This produces a higher secure key rate per pulse than conventional BB84 at most distances. The signal:decoy:vacuum probability ratios were 14:1:1, with relative pulse intensities of $\sim$1:0.3:0.00001. The mean photon number per pulse was set at 0.5. The system was operated using a random pattern of 10'000 symbols (10'000 ns) generated using the output of a continuous variable quantum random number generator \citep{Smith.2019}. Accumulated detection events were relayed back to the control PC and used to calculate raw bit rate and quantum bit error rate (QBER). To benchmark our photonic hardware, lower bound secure key rates were estimated for the asymptotic regime following the security proof of Ma \textit{et al.}\citep{Ma.2005}. For later ultra-long distance measurements we also use a finite key analysis to confirm positive secure bit rates in this case. 

The results of the experiment are shown in Figure 3. Both the forwards and backwards circuits of the transceivers were run simultaneously, with the performances shown separately. For the emulated loss measurements QKD performance was measured at 3 dB increments, with the attenuation varied by a VOA on the transmission channel. Combined bidirectional secure bit rate peaked at 3.07 Mbps for 7 dB loss. The highest attenuation to produce a secure key was 43 dB. QBER reached a minimum of 0.66\% at low attenuations. The increased QBER at the lowest attenuation (7 dB) was caused by increased SNSPD jitter when count rates were very high. Our high secure bit rates are achieved through the high repetition rate and exceptionally low QBER, and is particularly notable considering the relatively high loss of our actively modulated receiver circuit compared to state-of-the-art passive receiver circuits.

To substantiate the emulated loss measurements, we performed bidirectional QKD measurements over real fibre spools using lengths of 50 km and 180 km, the results of which are also shown in Figure 3. Both distances showed negligible deviation to the emulated loss curves, with 1.2 Mbps secure key rate in each direction over 50 km of fibre. The use of real fibre necessitated active stabilisation algorithms and dispersion compensation, details of which are given in Methods \ref{Timing}.

\subsection{Long-Term Stability and Resilience}

\begin{figure*}
\centering    
\includegraphics[width=1.0\textwidth]{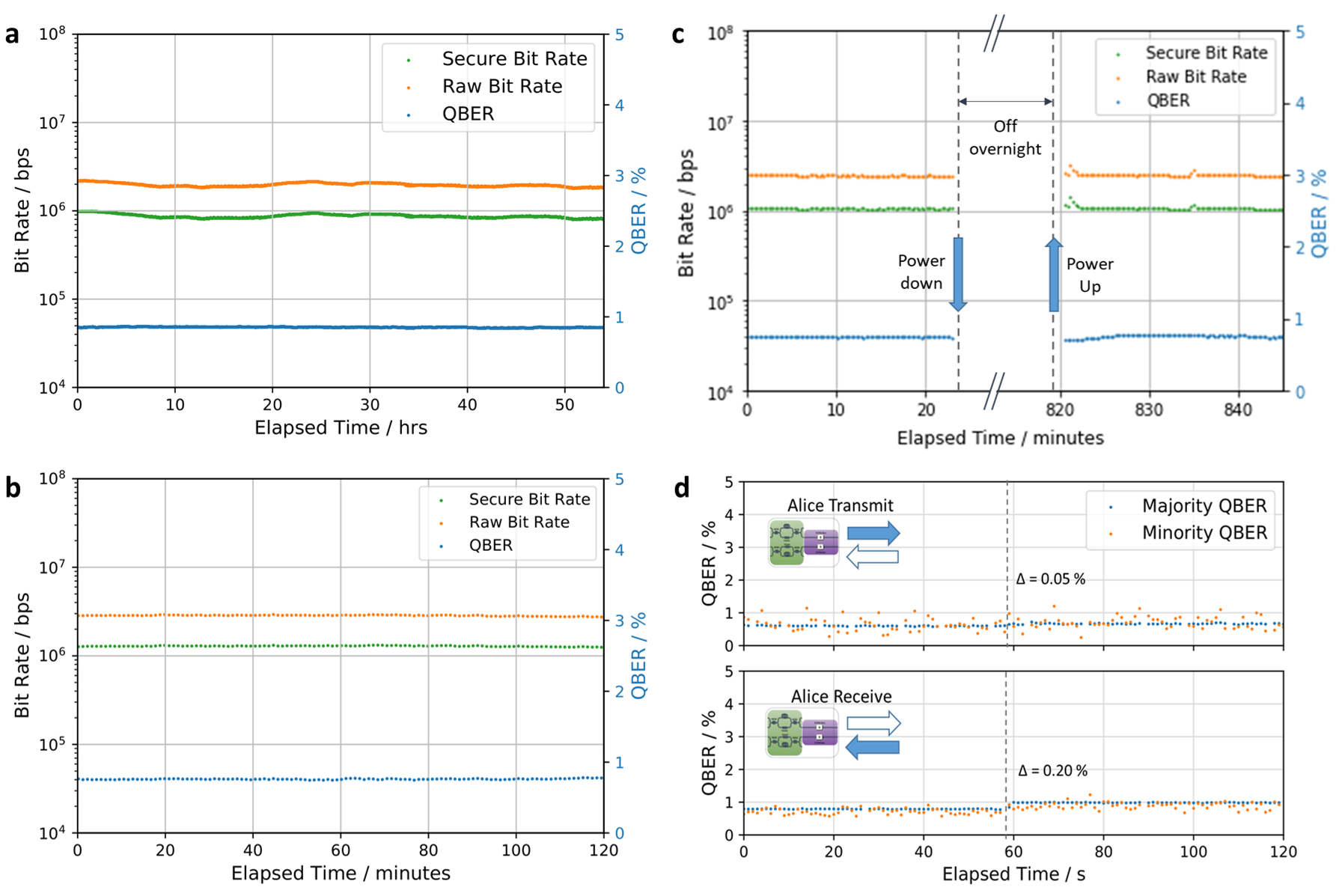}
\caption[SKR]{Demonstrations of chip stability. \textbf{(a)} 54 hours operation over 10 dB channel attenuation. \textbf{(b)} 2 hours operation over 50 km fibre spool channel. \textbf{(c)} Stability of the hybrid transceiver system when power-cycling over a 13 hour period. \textbf{(d)} Impact of running the complementary quantum communication channel on the majority and minority basis QBERs for forward (top) and backward (bottom) circuits. At 58s, the complementary circuit is powered-on causing a small increase in QBER due to predominantly electrical cross-talk between the EOPMs.}
\label{fig:Stability}
\end{figure*}

For a PIC with so many adjustable parameters, long-term stability is extremely desirable. Commonly, time-bin encoding QKD systems require three degrees of freedom to be stabilised: phase reference alignment between transmitter and receiver, timing (when do the pulses arrive at the receiver on a ps scale) and polarisation. Phase alignment will drift from temperature fluctuations at either interferometer, whilst the temporal and polarisation drift will come from extended lengths of fibre, such as over a real fibre spool transmission channel. We find that for our system the temperature stabilisation of the hybrid chips results in negligible interferometer phase drifts over any timescale we were able to practically measure. Thus, when operating over a short fibre link, where polarisation and timing do not significantly drift, we see stability over 54 hours with no stabilisation algorithms required (Figure 4.a.). In the case of real fibre spools our stability was limited by the gradual drifts of polarisation, since our setup did not have dynamic polarisation control though we could implement timing stabilisation through an algorithm further described in Methods \ref{Timing}. Over a 50 km real fibre spool we recorded 2 hours of stable operation (see Figure 4.b). Polarisation drift was also the likely cause of the small variations in count rate for the 10 dB channel data, though to a much smaller degree. Further, we also found that the phase stability of the chips remained constant through power-cycling. Figure 4.c demonstrates the stability of the system after 13 hours of downtime; the system returned rapidly to its previous set point with no re-optimisation required.

Figure 4.d illustrates the small, but measurable, increase in QBER due to the electrical cross-talk between the adjacent Tx and Rx EOPMs. We also saw optical cross-talk between the circuits, but measure there to be greater than 77 dB isolation.

\subsection{Long-Distance Unidirectional Operation}

Having demonstrated bidirectional QKD up to 180 km of real fibre, we wished to show the upper range limits of the system. For this we operated the hybrid chips in a unidirectional mode with only the 'Alice Transmit' circuits running. This helped with ultra-long distance operation in two ways: firstly by removing the increased QBER from electrical cross-talk in the EOPMs and, secondly, by removing the erroneous counts from optical cross-talk. At long distances the latter effect was more significant and we found that optical cross-talk photon counts were roughly on parity with the dark count rates of our detectors at 100 Hz.

Figure 5 shows the stability of the 'Alice transmit' system when operating over 250 km of real fibre spools. The system produced a QBER of 3.89\% and an estimated asymptotic secure key rate of 186 bps. Over the entire 2.5 hour period the system produced a block of 1.04x$10^7$ detection events and, following the finite key analysis proof of Lucamarini \textit{et al.} \citep{Lucamarini.2013}, we estimate that the block produces a secure key of 623 kb (67 bps). A summary of all key results presented thus far is given in Table 1. 

\begin{figure*}
\centering    
\includegraphics[width=0.7\textwidth]{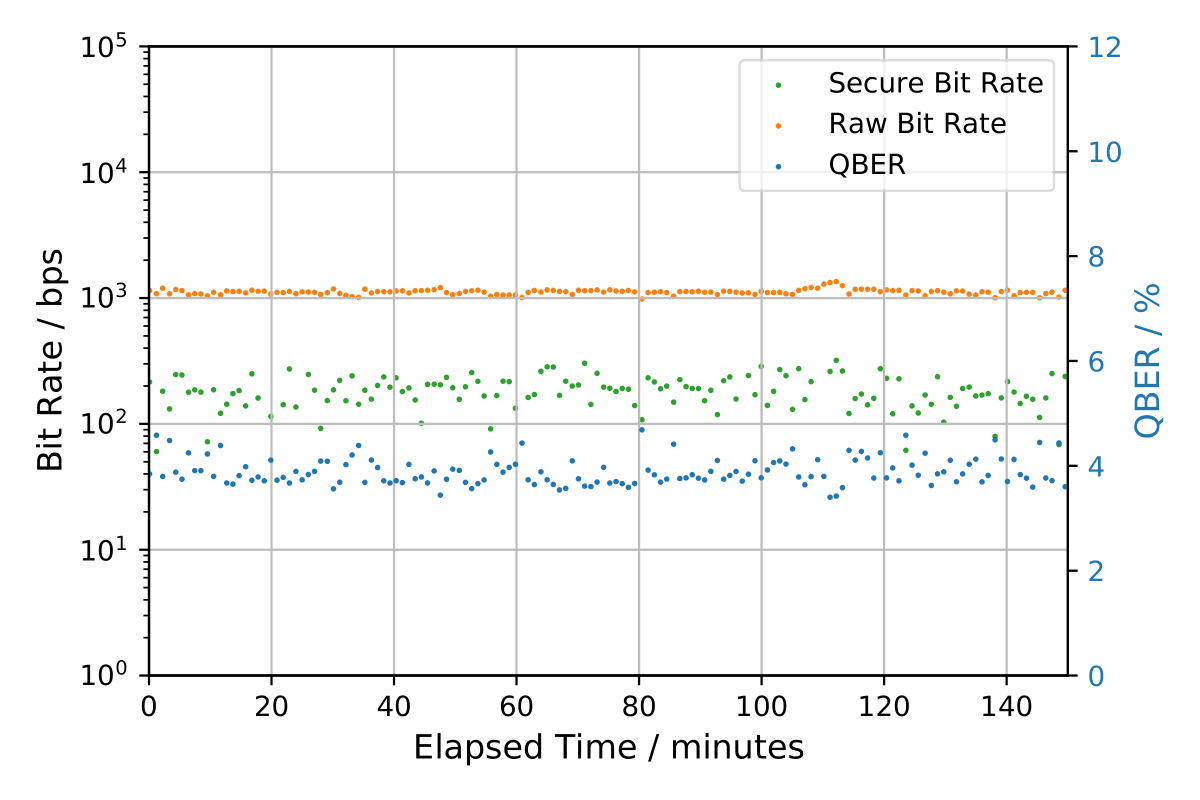}
\caption[SKR]{250km real fibre unidirectional operation. Raw bit rate, QBER and secure bit rate are shown averaged over 1 minute periods. Instantaneous secure bit rate is estimated in the asymptotic regime.}
\label{fig:250km}
\end{figure*}

\begin{table*}
\begin{center}
\begin{minipage}{\textwidth}
\caption{A summary of measured system performances. Measurements are taken during simultaneous bidirectional operation.}
\label{Table}
\begin{tabular*}{\textwidth}{@{\extracolsep{\fill}}lcccccc@{\extracolsep{\fill}}}
\toprule%
Channel & Circuit & QBER & Raw Bit Rate & SKR (Asymptotic)\\
\midrule
10 dB & Combined & 0.85\% & 4.25 Mbps & 1.82 Mbps \\
 & Forwards & 0.70\% & 2.09 Mbps & 904 kbps \\
 & Backwards & 0.99\% & 2.16 Mbps & 916 kbps \\
\midrule
50 km fibre & Combined & 0.89\% & 5.13 Mbps & 2.37 Mbps \\
(9.3 dB) & Forwards & 0.80\% & 2.61 Mbps & 1.21 Mbps \\
 & Backwards & 0.98\% & 2.52 Mbps & 1.16 Mbps \\
\midrule
180 km fibre & Combined & 1.11\% & 34.2 kbps & 14.1 kbps \\
(31.3 dB) & Forwards & 0.98\% & 17.3 kbps & 7.1 kbps \\
& Backwards & 1.23\% & 16.9 kbps & 6.9 kbps \\
\midrule
250 km fibre & Forwards & 3.89\% & 1.12 kbps & 186 bps \\
(44.0 dB) &  &  &  &  (67 bps finite-key)\\
\botrule
\end{tabular*}
\end{minipage}
\end{center}
\end{table*}

\section{Discussion}\label{sec12}

A QKD transceiver system has two advantages over independent transmitters and receivers. Firstly, it simplistically doubles the secure key rate over a given link. Whilst our system operates over two separate fibres in these experiments, it would be feasible to wavelength multiplex the two counter-propagating channels onto a single fibre. Secondly, as QKD research increasingly turns towards the development of larger networks, using transceiver nodes significantly increases the flexibility of a network. A transceiver can talk to any other transceiver (rather than communicating only with nodes of a complementary type) or can form a connection with two separate nodes, thereby flexibly forming structures such as ring networks.

When comparing our edge-coupling hybrid approach to conventional monolithic chips, our cost-benefit analysis must take into account the packaging costs of the edge-coupling process. The required multiple optical alignment steps add cost and complexity. However, we must also consider the fact that due to the realities of the material platforms, non-hybrid chip-based transceiver systems are likely to require at least two monolithic PICs per node. With a hybrid transceiver PIC we can reduce this down to a single chip assembly, whilst allowing both transmitter and receiver circuits to access the best properties from two different material platforms. 

Similarly, we can compare our edge-coupling hybrid approach to alternative hybrid or heterogeneous solutions, such as flip-chip or photonic wire bonding. For our purposes we believe that edge-coupling currently has two key strengths: low optical coupling losses and sufficiently mature manufacturing that is accessible in low volumes. As parts of our chip are used as a receiver, the optical loss is critical. The fact that we find this can be repeatably achieved with currently available commercial manufacturing techniques is important and should be of interest to those working on other applications of quantum photonic integration. However, we do not discount the potential advantages of alternative hybrid and heterogeneous integration approaches. For large circuits, as will likely be required for photonic quantum computing for example, the inherent scalability of some of the alternatives becomes a significant advantage. Indeed, even for smaller circuits, such as used in QKD, wafer-scale approaches may prove to have more economical manufacturing. Recent work has shown Si-InP photonic wire bonding coupling losses as low as 0.4 dB \citep{Billah.2018} and repeatable sub-1dB losses resilient to thermal cycling \citep{Blaicher.2020}.The ongoing development of various hybrid and heterogeneous integration methods and their permeation into industry offers exciting avenues for future work.

In comparison to previous on-chip time-bin QKD demonstrations, we find the low minimum QBER of the system notable. Sub-1\% QBER, as we measure, is more commonly associated with time-of-arrival measurements due to the higher extinction ratio \citep{Sibson.2017,Boaron.2018}. In contrast, qubits encoded in differential phase, like in the present work, are more vulnerable to intensity imbalance and delay-line mismatch between Alice and Bob. Here, we find that the precise manufacturability of SiN yields excellently matched interferometers, which produces high contrast interference. Further, the availability of heaters on SiN allows the AMZI structures to be finely adjusted and offset manufacturing errors further. This stands favourably in comparison to other material platforms used for QKD receivers, such as planar light-wave circuits on silica. 

We can also compare alternative material platforms for the modulator component. For example, thin film lithium-niobate-on-insulator (LNOI) is emerging as a powerful platform for on-chip modulation, with an exceptionally strong electro-optic effect and low propagation losses \citep{DiZhu.2021}. However, InP still represents the most mature platform for this purpose and allows for convenient integration of lasers.

From a practical perspective, our devices greatly benefit from the low voltage and power requirements of the EOPMs. Once modulation voltages fall below 5 V, the design of efficient drive electronics using CMOS compatible components becomes more practical. Additionally, a lower modulation voltage drastically reduces the power consumption of modulation, which can allow for commercial systems with more sustainable power requirements. The low V\textsubscript{$\pi$} of 4 V demonstrated here compares well to previous silicon-based QKD modulators \added{\citep{Sibson.2017b,Geng.2019,Wei.2020,Zhu.2022}, whilst the strong phase modulation efficiency of the EOPMs allows this to be achieved with a compact modulator that keeps optical losses low.} 

The stability of the chips under reasonable temperature stabilisation is also a prerequisite for practical implementation. Our results show that the interferometers remain phase stable over long periods of time (50 hours) and through power cycling of the electronic systems. Further, we have shown that the system can operate over real fibre spools given appropriate stabilisation against temporal and polarisation drifts. Notably, recent efforts have shown that polarisation sensitivity can be removed when using low birefringence silica receivers \citep{Sax.2023}.

Further work could focus on real-time system operation. As an option to reduce the footprint further, the externally modulated laser chip could be brought onto the InP section of the hybrid. This does not present any fundamental challenge, since both are based on InP, but will have engineering and manufacturing implications to be considered. We note that in some circumstances have a separate laser source can be advantageous, for example when matching a QKD system to the available wavelength channels on an existing fibre. With the footprint as it stands the hybrid and external laser PICs could be incorporated into pluggable modules. 

Taken together, the above results and discussion point towards the fact that hybrid integration can be used to produce a QKD chip with state-of-the-art performance that is stable, manufacturable, and that has reasonable control and power requirements. More broadly, this supports the potential of hybrid integrated photonics to improve the performance of quantum photonic devices beyond current material platform limitations.

\newpage

\begin{small}

\section{Methods}

\subsection{Insertion Loss Measurements}\label{OLM}

Optical loss measurements were carried out using 1550 nm polarised light with the axis of polarisation rotated to align with the transverse-electric mode of the SiN PICs. A variety of different paths through the PICs were measured in order to estimate individual component contributions to the insertion loss. In order to estimate delay line losses, the TOPMs in the AMZI structure were adjusted so as to send all light down the short arm and then all the light down the long arm, the loss difference giving a delay line loss. Each hybrid PIC contains two fibre-SiN and two SiN-InP interfaces. The interface loss values given in Figure 2 are thus double the quoted losses per interface. When giving estimated interface losses in Figure 2, we work off the basis of the average of measurable interfaces on the same hybrid chip. The values given in Figure 2 should be taken as best guess estimates. For circuits 3-A and 4-B, we cannot be sure that the increased insertion loss comes from the InP, rather than the SiN-InP interfaces, but we believe it to be the most likely case. Raw data is given in Supplementary Table 1. 

\subsection{Laser Source PICs}

For our EMLs we used custom-made indium phosphide PICs containing a distributed feedback (DFB) laser and an electro-absorption modulator (EAM). The DFB laser was gain-switched with a 1/3 duty cycle at 1 GHz to produce short optical pulses of width $\sim$100 ps. The lasers were designed to operate at roughly 1550nm, with one laser PIC operating at centre wavelength 1550.20 nm and the other at 1553.64 nm. The output of the laser PICs was filtered down to width 0.1 nm using external filters, which was found to improve the QBER of the system. Our decoy-state protocol requires us to produce three different intensity levels: 'signal', 'decoy' and 'vacuum'. Gain-switching the laser without attenuation produced signal pulses. Applying a 500 ps $\sim$1 V pulse to the EAM attenuated the pulses to decoy intensity. To produce vacuum pulses the gain-switching pulse was skipped, thereby leading to no emission. The laser drive and EAM signals were produced by the marker outputs of an AWG operating at 12 GS/s.

\subsection{Hybrid PIC Control Signals}

Each hybrid transceiver PIC required five DC heater channels, two DC EOPM bias channels and two four-level RF EOPM control signals. The DC controls were set by precision source-measure units with a resolution of 0.0002. The necessary heaters voltages ranged from 0 - 20V, whilst all EOPM biases were set to +2.5V. The bias voltage for the EOPMs were set so as to prevent spontaneous emission from the EOPM, which was found to cause spurious detection events at the SNSPDs. The four-level RF control signals were produced using AWGs with a 1/3 duty-cycle square wave at 1 GHz and at 12 GS/s. The RF signals were amplified using amplifiers so as to bring the V\textsubscript{$\pi$} to $\sim$4 V. 

\subsection{Photon Detection}

For our external detectors, we used superconducting nano-wire single photon detectors (SNSPDs) cooled down to 2 K using liquid helium cryogenically. Each detector operated at $\sim$85\% detection efficiency with dark counts of around 100 Hz. The detectors were operated in a free-running mode. However, a 300 ps gate applied in software post-processing was used to select only the photons corresponding to the desired interference events. 

\subsection{Timing and Fibre Stabilisation}\label{Timing}

For each QKD 'direction', our system contained five control signals: laser pulsing, EAM, Alice EOPM, Bob EOPM and detector trigger. All five signals for each direction were controlled by a single AWG to maintain easy synchronisation and allow for relative delay adjustment, resulting in two AWGs with completely independent timing, one for each direction. Delays were adjusted in 10 ps increments so as to minimise the system QBER. Once set, these delays were found to be stable. 

The exception to this stability was when using real fibre spools, whereby the relative delay of the Bob drifted significantly compared to Alice due to thermal expansion and contraction of the fibre spool. For this case, a stabilisation algorithm was devised to compensate for fibre drift. Through this algorithm, the PC was able to 'lock on' to the detected pulses from the SNSPDs and dynamically compensate for delay changes from the fibre spool for the Bob EOPM and detector triggers. When operating over fibre spools we also added appropriate dispersion compensation to the transmitter systems. 

\subsection{Secure Key Estimation}

When estimating secure key production in the asymptotic regime, we use a forward error correction efficiency factor of 1.25. When estimating secure key production in the finite block regime, we use a forward error correction efficiency factor of 1.25 and a total security parameter of $10^{-10}$.

\end{small}

\backmatter

\section*{Declarations}

\bmhead{Data Availability}

The datasets generated during and/or analysed during the current study are available from the corresponding author upon reasonable request.

\bmhead{Code Availability}

The data analysis software code used during the current study is available from the corresponding author upon reasonable request.

\bmhead{Acknowledgments}

We thank T. Roger for fruitful discussions and P. R. Smith for allowing us access to their quantum random number datasets. J.A.D. thanks R.V. Penty for his guidance and supervision. J.A.D. acknowledges funding from the UK's Engineering and Physical Sciences Research Council under the Industrial Cooperative Awards in Science \& Technology (CASE) programme.

\bmhead{Authors' Contributions}

T.K.P., J.A.D. and A.J.S. devised the experiments. T.K.P. designed the photonic integrated circuits. J.A.D., T.K.P. and H.D. assembled the experimental setup. J.A.D. characterised the photonic integrated circuits, wrote the control software, acquired the system performance data and carried out the data analysis. R.I.W. and D.G.M. contributed finite-key analysis code. J.A.D. wrote the manuscript with contributions from all authors. T.K.P. and A.J.S. supervised the project.

\bmhead{Competing Interests}

The authors declare no competing financial or non-financial interests.

\bibliography{references}

\end{document}